# Laboratory study of antenna signals generated by dust impacts on spacecraft

Mitchell M. Shen[1,2], Zoltan Sternovsky[1,2], Mihály Horányi[1,3], Hsiang-Wen Hsu[1], and David Malaspina[1,4]

1. *Laboratory for Atmospheric and Space Physics, University of Colorado, Boulder, CO 80303*
2. *Smead Aerospace Engineering Sciences Department, University of Colorado, Boulder, CO 80303*
3. *Physics Department, University of Colorado, Boulder, CO 80309*
4. *Astrophysical & Planetary Sciences Department, University of Colorado, Boulder, CO 80309*

***Corresponding author: Mitchell.Shen@lasp.colorado.edu***

Key Points:

(1) A simple model based on induced charging is shown to reproduce dust impact signals detected by antenna instruments.

(2) Key parameters of the dust impact plasma are obtained from fitting the model to data from laboratory measurements.

(3) The presented model is applicable to a range of space missions with antenna instruments.

Key Words: cosmic dust, dust detection, antenna instruments

# Abstract

Space missions often carry antenna instruments that are sensitive to dust impacts, however, the understanding of signal generation mechanisms remained incomplete. A signal generation model in an analytical form is presented that provides a good agreement with laboratory measurements. The model is based on the direct and induced charging of the spacecraft from the collected and escaping fraction of free charges from the impact generated plasma cloud. A set of laboratory experiments is performed using a 20:1 scaled down model of the Cassini spacecraft in a dust accelerator facility. The results show that impact plasmas can be modeled as a plume of ions streaming away from the impact location and a cloud of isotropically expanding electrons. The fitting of the model to the collected antenna waveforms provide some of the key parameters of the impact plasma. The model also shows that the amplitudes of the impact signals can be significantly reduced in typical space environments due the discharging effects in the ambient plasma.

# 1. Introduction

Antenna instruments are designed to measure electric fields and/or the electromagnetic waves propagating in planetary ionospheres and magnetospheres, or the interplanetary medium. Antenna instruments, however, are also known to be sensitive to the impacts of cosmic dust particles as demonstrated by several missions, including Voyager 1 & 2, Cassini, STEREO (Solar TErrestrial RElations Observatory), Wind, MMS (Magnetospheric Multiscale), or the Parker Solar Probe [*Gurnett et al.*, 1983, 1997, 2004, 2005; *Aubier et al.,* 1983; *Meyer-Vernet et al.,* 1986, 2009, 2017; *Kurth et al.,* 2006*; Wang et al.,* 2006*; Morooka et al.,* 2011*; Hill et al.,* 2012*; Ye et al.* 2014, 2016a, 2016b, 2018, 2019; *Zaslavsky et al.,* 2012; *Malaspina et al.,* 2014, 2020; *Thayer et al.,* 2016; *Kellogg et al.,* 2016; *Vaverka et al.,* 2017, 2019; *Szalay et al.*, 2020]. Antenna instruments register the dust particles through the transient plasma cloud generated upon impact on the spacecraft or the antennas, and the associated charge separation and the collection of charges [*Oberc*, 1996].

The quantitative understanding of the generation of dust impact related waveforms remained incomplete. Several models have been used for the analysis of data from a range of missions and considering the configuration of the antennas (dipole vs. monopole). These models have evolved over time, starting with the simple charging of the antenna by collecting electrons from the impact plasma [e.g., *Gurnett et al.*, 1983], or the generation of the 'polarization' electric field from charge separation of the expanding impact plasma [*Wang et al.*, 2006], to the recent and comprehensive models based on the physics of expanding plasma clouds [*Oberc*, 1996; *Meyer-Vernet et al.,* 2014, 2017]. The laboratory simulation of antenna coupling mechanisms became feasible by the commissioning of the dust accelerator facility at the University of Colorado [*Shu et al.*, 2012]. *Collette et al.* [2015] have identified three mechanisms that can lead to the generation of impact signals: (1) spacecraft charging, where some fraction of the impact plasma is recollected, and the polarity of the signal is affected by the bias potential of the spacecraft, (2) antenna charging for impacts occurring near the antenna base, and (3) induced charging, where the space charge of the expanding electrons and ions of the plasma cloud are detected. *Nouzák et al.* [2018] refined these measurements by using a scaled-down model of the Cassini spacecraft for the measurements. These measurements characterized the variations of the amplitude and the polarity of the antenna signals due to applied bias potentials and the configuration of the antennas. This study was expanded to include the effects of magnetic fields by *Nouzák et al.* [2020]. Combined, the laboratory studies provided the qualitative understanding of how the collected and escaping fractions of the electrons and ions from the impact plasma can lead to signal generation, either through the charging of the spacecraft, or the induced effects of the escaping charged particles. In a separate study, *O'Shea et al.* [2017] showed that that charging of the antenna by the impact plasma expanding away from the spacecraft is of secondary importance in comparison, unless the dust impact occurs close to the base of the sensing antennas. In addition, analysis of the laboratory measurements allows the determination of some of the relevant parameters of the impact plasma, including the effective temperatures of the electrons and ions. The effective temperatures of impact plasmas have also been studied separately in laboratory conditions for a small number of dust-target material combinations [*Collette et al.*, 2016; *Kočiščák et al.*, 2020].

This article presents the first quantitative analytical model describing the mechanisms how antenna signals are generated under a few simplifying assumptions. It is shown that the model can be fitted accurately to waveforms measured in the laboratory. The overarching motivation is the ability of calculating the total impact charge from the antenna waveforms, as the impact charge is the measure of the dust particle's mass and impact speed, as the three quantities can be related through calibration measurements [e.g., *Collette et al.*, 2014].

## 2. Experimental Setup

The measurements are performed using the reduced-size model of the Cassini spacecraft described in detail by *Nouzák et al.* [2018]. Briefly, the simplified 20:1 scaled model consists of the cylindrical body, the High Gain Antenna (HGA), the magnetometer boom (MAG), and the three orthogonal antennas, $E_U$, $E_V$, and $E_W$ (Fig. 1). The Al body is coated with graphite paint that provides a uniform surface potential [*Robertson et al.*, 2004]. The antennas are made from stainless steel rods. The dust beam is pointed at a 25×25 mm$^2$ tungsten (W) foil that is mounted near the edge of the HGA, as shown in Fig. 1. The antennas and the W target are cleaned using organic solvents. The model is mounted in the middle of a large experimental vacuum chamber using electrically isolating brackets. The chamber (1.2 m in diameter and 1.5 m long) is evacuated to < 10$^{-6}$ Torr for the measurements without ambient plasma. The antennas are relatively far from the location of dust impacts, and thus the expanding impact plasma. In this arrangement, the dominant signal generation mechanism is from the charging of the spacecraft body.

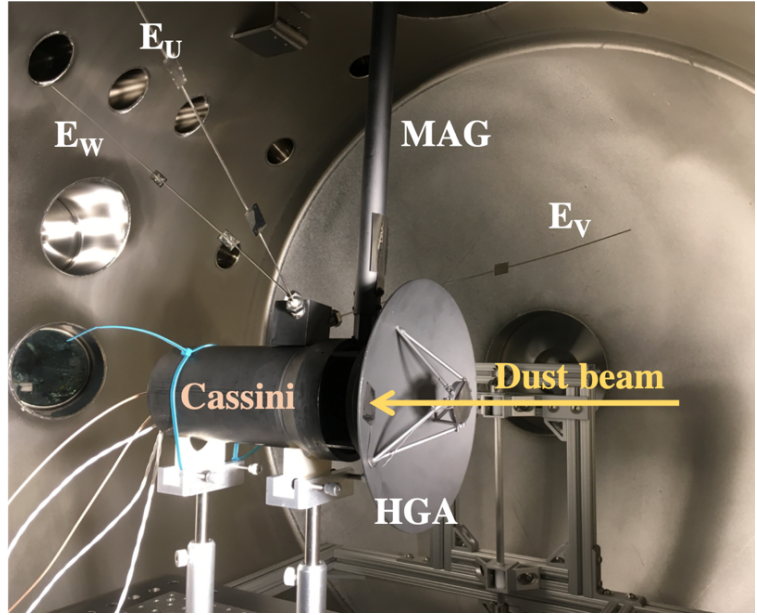

**Fig. 1:** The model spacecraft mounted inside the vacuum chamber. The dust beam is pointed at the W foil mounted near the edge of High Gain Antenna (HGA) as indicated by the arrow. The diameter of the HGA is approximately 20 cm.

Iron (Fe) dust particles are used in the experiments. The spherical particles are typically 40 – 50 nm in radius and are accelerated to velocities of 20 – 25 km/s using the accelerator facility at the University of Colorado [*Shu et al.*, 2012]. The dust-target (Fe-W) combination is selected such that the results are directly comparable to previous studies [*Collette et al.*, 2015; 2016; *Nouzák et al.*, 2018; 2020]. The impact speed range is limited to keep the basic parameters of the impact plasma comparable over the whole set of the measurements.

The front-end electronics is integrated into the cylindrical part of the spacecraft model (SC). It measures the voltage differences between the $E_U - E_V$ antennas in a dipole mode, and the $E_w - SC$ in a monopole mode (Fig. 1). The measurements and modeling in this article are limited to the latter. The previous measurements reported by *Nouzák et al.* [2018] indicated that some of the fine structures in the antenna signals were not fully resolved due to the limited bandwidth of the previous setup (50 Hz – 400 kHz). The bandwidth of the electronics has been extended to 270 Hz – 5 MHz, which resolves this issue. The upper limit of the bandwidth corresponds to a rise time of about 70 ns, and the slew rate of the amplifiers is 35 V/μs, which does not introduce further limitations. The voltage gain of the amplifier is $G = 50$. The biasing scheme of the antennas and the SC remained the same, where electric potentials can be applied to the antennas and the SC independently through a resistor with an effective value of $R_{BIAS} = 5\ M\Omega$. In the measurements reported here, the same bias potential is applied to all components. The data are recorded using a fast digitizing oscilloscope at a sampling rate of 200 MS/s.

## 3. Antenna Signal Generation Model

### 3.1. Model Description

It is assumed that the dust particle generates an impact ionization plasma cloud consisting of free electrons and cations (anions are neglected). The total impact charge ($Q_{IMP}$) scales with the mass ($m$), and impact velocity ($v$) of the particle following a power law in a form of:

$$Q_{IMP} = Q_i = |Q_e| = \gamma m^\alpha v^\beta, \tag{1}$$

where parameters $\gamma, \alpha$, and $\beta$ can be determined from laboratory calibration measurements [e.g., *Auer,* 2001; *Collette et al.,* 2014]. An assumption is made here that the total charges of positive ($Q_i$) and negative ($Q_e$) charge carriers are equal. The next necessary assumption is that $Q_{IMP}$ is reasonably small such that (a) it introduces only a relatively small perturbation to the equilibrium potential of the spacecraft ($V_{SC}$), and (b) the electrons and ions in the impact plasma decouple from one another (meaning that the electrons are no longer trapped by the positive space charge of the ion cloud) over a distance that is small compared to the characteristic size of the spacecraft ($R_{SC}$). The corresponding relations are $Q_{IMP} < C_{SC}V_{SC}$, and $Q_{IMP} < 4\pi\varepsilon_0 \frac{T_e}{e_0} R_{SC}$, respectively, where $C_{SC}$ is the capacitance of the SC, $T_e$ is the temperature of the electrons of the impact plasma (expressed in the units of eV), $\varepsilon_0$ is the permittivity of free space, and $e_0$ is the elementary charge. The faster electrons decouple from the plasma soon as the space potential of the ion cloud drops below the electron temperature during the expansion. The corresponding assumption from above was derived by *Meyer-Vernet et al.* [2009] and *O'Shea et al.* [2017], and allows treating the ions and electrons independently, and as particles whose trajectories are mostly governed by the potential of the SC. For a typical SC operating near 1 AU in interplanetary space [e.g., *Zaslavsky*, 2015], and $T_e = 5$ eV, the above conditions are satisfied for impact charges $Q_{IMP} < 5 \times 10^{-10}$ C.

The total impact charge, both the positive ion (*i*) and electron (*e*) parts, can be divided into fractions that are collected by the SC, and escaping from the SC, i.e.:

$$Q_i = Q_{i,esc} + Q_{i,col}$$
$$Q_e = Q_{e,esc} + Q_{e,col}. \qquad (2)$$

The collected and escaping fractions are determined by their effective temperatures ($T_e$ and $T_i$), and the spacecraft potential. For $V_{SC} > 0$,

$$Q_{e,esc} = -\kappa\, Q_{IMP}\, e^{\frac{-V_{SC}}{T_e}}$$
$$Q_{e,col} = -Q_{IMP} - Q_{e,esc}$$
$$Q_{i,esc} = Q_{IMP}$$
$$Q_{i,col} = Q_{IMP} - Q_{i,esc} = 0. \qquad (3)$$

The third equation from above expresses the assumption that ions are moving away from the impact location in form of a plume. On the other hand, the light and fast electrons acquire an isotropic velocity distribution through collisions during the early phases of the expansion. The geometric coefficient $\kappa$ (with a value between 0 and 1) is introduced in order to account for the geometry of the SC body around the impact site. In an ideal case of a semi-infinite target and no obstructions in the field-of-view of the expansion, $\kappa = 1/2$ would account for the fact that half of the electrons are moving away from the SC and will be escaping, provided that it is energetically feasible. For a negative spacecraft potential ($V_{SC} < 0$) the corresponding equations are:

$$Q_{e,esc} = -\kappa\, Q_{IMP}$$
$$Q_{e,col} = -Q_{IMP} - Q_{e,esc}$$
$$Q_{i,esc} = Q_{IMP}\, e^{\frac{V_{SC}}{T_i}}$$
$$Q_{i,col} = Q_{IMP} - Q_{i,esc}. \qquad (4)$$

The effective temperatures of the positive and negative charge carriers have been measured in laboratory conditions using a dust accelerator. The measurements by *Collette et al.* [2014], using

Fe dust particles impacting onto a W target found that the effective temperatures of negative and positive charge carriers vary from $T_e = 1$ eV to 4 eV, and $T_i = 4$ eV to 20 eV, respectively, with increasing impact speed from 4 km/s to 20 km/s. *Nouzák et al.* [2018] found $T_e$ to be on the order of 1 eV, and $T_i$ between 10 – 15 eV for impact speed in the range of 20–25 km/s using the same Fe/W dust-target material combination. In a more recent study by *Kočiščák et al.* [2020], using an organic-coated olivine dust sample impacting onto a W target, the authors found $T_i \approx 7$ eV over a wide impact speed range of 2 – 18 km/s, while $T_e$ varied between 1 – 10 eV non-monotonically over the same impact velocity range. The authors suggested that the varying relative fraction of anions to free electrons is the reason for the significant variation observed for $T_e$. Olivine was selected for the study as a rock-forming mineral that is also common in meteorites, and the organic coating on the dust sample was applied in order to make the sample usable in the accelerator [*Fielding et al.*, 2015].

The next step is to describe the temporal evolution of the antenna signal. We will assume the simple case of a monopole configuration, where the antenna is far from the impact location and not in the way of the expanding plasma cloud. The antenna instrument measures the potential of the antenna with respect to the SC body. In space, both the antenna and the SC are at the equilibrium floating potentials, where the charging currents are in balance. The two dominant currents in interplanetary space are due to photoelectron emission and the collection of electrons from the ambient plasma [*Zaslavsky,* 2015], which result in floating potentials on the order of +5 V. Since the electronics is AC-coupled, the monopole antenna measures the deviation in the potential differences due to the transient dust impact event, i.e.:

$$V_{meas}(t) = \delta V_{ANT}(t) - \delta V_{SC}(t). \tag{5}$$

We will further assume a dust impact on the spacecraft body and that the antenna potential is not affected (see configuration shown in Sec. 2). In other words, the measured signal is simply the variation of the spacecraft potential, $V_{meas}(t) \cong -\delta V_{SC}(t)$. The spacecraft potential is related to charge on the spacecraft through its capacitance, i.e. $\delta V_{SC}(t) = \delta Q_{SC}(t)/C_{SC}$. For completeness it is added that there is some capacitive coupling between the antenna and the SC. The consequence of such coupling is that the measured voltage is somewhat reduced, i.e. $V_{meas}(t) \cong -\Gamma \delta Q_{SC}(t)/C_{SC}$, where $\Gamma$ is known as a coupling parameter with a value between 0 – 1. The extent of the capacitive coupling, however, has not been established experimentally for the setup, and for simplicity $\Gamma = 1$ is assumed.

The qualitative description of the spacecraft charging model is based on the works by *Meyer-Vernet et al.* [2017] and *Nouzák et al.* [2018; 2020], and its high-level description is also provided in *Ye et al.* [2019] and *Mann et al.* [2019]. Briefly, a fraction of the electrons from the impact plasma (described in Eqs. (3) or (4)) escape quickly, resulting in a rapid positive charging of the spacecraft. The escape of the slower ions takes place simultaneously and it is charging the spacecraft negatively. Once the escape processes are completed, the spacecraft is left with the net collected charge of electrons and ions from the impact plasma. Throughout the process, the effect of the ambient plasma drives the spacecraft potential back to equilibrium. The following charging equation describes the transient event of a dust impact (occurring at time $t = 0$):

$$\delta Q_{SC}(t) = Q_{i,col} + Q_{e,col} + Q_{i,esc} \frac{R_{SC}}{R_{SC}+v_i t} + Q_{e,esc} \frac{R_{SC}}{R_{SC}+v_e t} + \int_0^t I_{dis}(\tau)d\tau \quad (6)$$

This relation assumes that the generation of the impact charge is an instantaneous process. It is important to note that a small cloud of charge in the close vicinity of the SC has a similar effect on its potential as the same charge collected on the SC. This is called induced charging. For this reason, the collection of the electron and ion charges by the SC ($Q_{e,col}$ and $Q_{i,col}$) can be considered to take place instantaneously, even though some of the electrons or ions are on ballistic orbits before returning to SC. For a spherical approximation of the spacecraft body (effective radius $R_{SC}$), the induced charges of the escaping fractions ($Q_{e,esc}$ and $Q_{i,esc}$) scale as $\sim 1/r$ with radial distance [*Jackson*, 1999], as expressed in the third and fourth terms on the right-hand-side of Eq. (6). The characteristic expansion speeds of ions and electrons are $v_i$ and $v_e$, respectively. The last term in Eq. (6) accounts for discharging through the ambient plasma environment. *Zaslavsky* [2015] and *O'Shea et al.* [2017] derived the equations and the time constant of the discharge current for conditions, where the equilibrium spacecraft potential is maintained by the balance between photoelectron emission and electron collection from the ambient plasma environment (e.g., in interplanetary space). The corresponding discharge time-constant for an assumed spherical symmetry is

$$\tau_{dis} = \frac{1}{2}\frac{C_{SC}T_{ph}}{e_0 n_e w_e S_{SC}}, \tag{7}$$

where $T_{ph}$ is the characteristic temperature of photoelectrons, $n_e$ is the density of ambient plasma, $S_{SC}$ is the surface area of the spacecraft, and $w_e = \sqrt{e_0 T_{e,ap}/2\pi m_e}$ is the electron thermal speed, with $T_{e,ap}$ and $m_e$ being the temperature of the ambient plasma electrons and electron mass, respectively.

Differentiating Eq. (6) and using Eq. (2) we arrive to the transient charging equation in the following form:

$$\frac{d}{dt}(\delta Q_{SC}) = -Q_{i,esc}\frac{v_i R_{SC}}{(R_{SC}+v_i t)^2} - Q_{e,esc}\frac{v_e R_{SC}}{(R_{SC}+v_e t)^2} - \frac{Q_{SC}}{\tau_{dis}}. \tag{8}$$

This equation can be solved numerically.

## 3.2. Antenna Signals in Interplanetary Space

In this section we illustrate the solution of the signal-generation model from above for an ideal spacecraft in interplanetary space. Specifically, we will consider the STEREO spacecraft at 1 AU and refer to the works by *Zaslavsky* [2015] and *O'Shea et al.* [2017]. The relevant parameters considered are the following: The equilibrium floating potential of the spacecraft is $V_{SC} = 5.5$ V, the spacecraft capacitance is $C_{SC} = 200$ pF, the surface area is $S_{SC} = 10$ m², with an effective radius of $R_{SC} = 1$ m. The typical parameters of the plasma environment are $n_e = 5$ cm⁻³, $T_{e,ap} = 8$ eV, and $T_{ph} = 2$ eV. For these condition $w_e = 0.47 \times 10^6$ m/s and $\tau_{dis} = 0.053$ ms. The following parameters are used to describe the impact plasma: $\kappa = 0.5$, $v_e = 10^6$ m/s, $v_i = 10^4$ m/s, $T_e = 7$ eV and $T_i = 25$ eV. The latter values will be justified in Sec. 4.

Figure 2 shows the antenna signal for a model spacecraft operating in interplanetary space near 1 AU. For simplicity, the impact charge is set to $Q_{IMP} = C_{SC}(1\,V) = 200$ pC. The sharp negative going feature is known as the 'preshoot' and is generated by the rapid escape of free electrons from impact plasma, as it has been explained by *Nouzák et al.* [2018; 2020]. The signal shape and characteristic time-constants are in good qualitative agreement with the 'triple-hit' signals from the STEREO spacecraft described by *Zaslavsky et al.* [2012], *Collette et al.* [2015], or *O'Shea et al.* [2017].

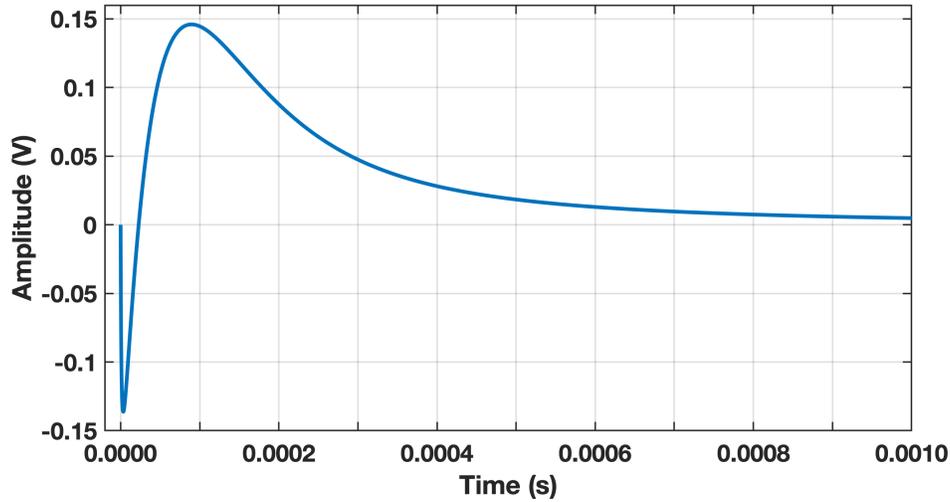

**Fig. 2:** The antenna signal for a model spacecraft operating in interplanetary space near 1 AU. See text for more detail.

A noteworthy observation in Fig. 2 is that the amplitude of the signal is significantly lower than the $Q_{IMP}/C_{SC} = 1$ V value. Following Eq. (3) and the parameters defined above, the escaping electron charge is $-0.23 \times Q_{IMP}$. However, the amplitude of the preshoot never reaches the full 0.23 V because the ions from the impact plasma are expanding simultaneously, albeit with a slower speed. Without discharge from the ambient plasma environment, the net collected charge on the SC after the plasma expansion is completed is $Q_{col} = Q_{e,col} + Q_{i,col} = -(Q_{e,esc} + Q_{i,esc}) = -0.77 \times Q_{IMP}$. Such collected charge would generate a signal with corresponding amplitude of 0.77 V. The actual amplitude, however, is significantly lower, only about 0.15 V. The reason for this is the ongoing discharge from the ambient plasma, where the corresponding time-constant from Eq. (7) is relatively short. Figure 3 shows the details of contributions from the escape of electrons ($V_{e,esc}(t)$) and ions ($V_{i,esc}(t)$) along with the measured voltage accounting for ambient discharging ($V_{meas}(t)$).

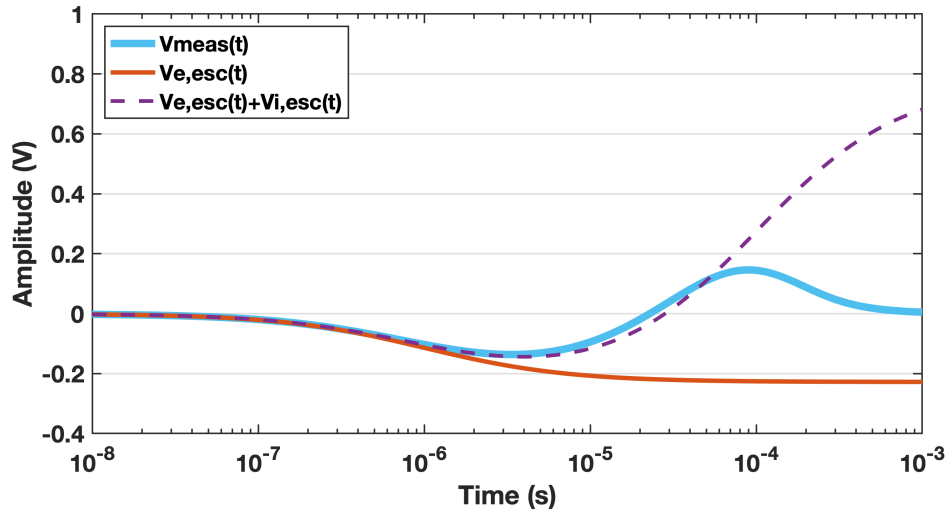

**Fig. 3:** The antenna signal and contributions from selected terms on a logarithmic time scale. $V_{meas}(t)$ is the same as the measured signal shown in Fig. 2. The red curve is the contribution from the escaping electrons, $V_{e,esc}(t)$ and the purple dashed curve represents the sum of signals from the escape of electrons and ions, $V_{e,esc}(t) + V_{i,esc}(t)$. The detailed list of parameters used for the calculations is shown in Table. I.

In summary, the antenna signal model from Sec. (3.1) has three characteristic time constants: (1) electron escape is described by $\tau_{e,esc} = R_{SC}/v_e$, (2) ion escape by $\tau_{i,esc} = R_{SC}/v_i$, and (3) the plasma discharge time constant $\tau_{dis}$. The amplitudes of the preshoot and the main signal are affected by the interplay between these competing effects. For the model case of a spacecraft in interplanetary space described above, the signal amplitude is significantly reduced compared to the theoretical maximum of $Q_{IMP}/C_{SC}$.

### **3.3.** Antenna Signals Measured in the Laboratory

This section discusses the similarities and differences between the measurements in a controlled laboratory environment and those that can be expected in space. The discharge of the SC through the ambient plasma is modeled as a discharge through the biasing resistor ($R_{BIAS}$), as described by *Nouzák et al.* [2018]. The corresponding discharge time constant, $R_{BIAS}C_{SC}$, is selected to be relatively long, such that the amplitudes of the signals are not significantly affected (Fig. 4). Other than discharging, the ambient plasma has little effect on the expansion of the impact plasma and related charging processes. Provided that near 1 AU the Debye length of that ambient plasma is large compared to the characteristic size of the spacecraft, the potential profile around the spacecraft is not altered significantly by the plasma sheath.

The measurements and data analysis are limited to impact speeds ≥ 20 km/s, similarly to the work done by *Nouzák et al.* [2020]. The goal of this article is to demonstrate a simplified physical model from Sec. (3.1) can be used to explain the generated signal shapes. Generally, the parameters of the impact plasma, including ion composition, or the effective temperatures of positive and negative charge carriers, vary with impact speed. In particular, for ≥ 20 km/s volume ionization is expected to dominate and the plasma consists mainly of electrons and atomic cations [*Mocker et al.* 2013, *Hillier et al.,* 2014]. In other words, the characteristics of the impact plasmas from the described measurements are the closest to assumptions made for the signal generation model described above.

The laboratory measurements are performed using an Fe-W target-dust material combination for reasons described in Section 2. The parameters determined from the experimental data described below may or may not be representative to cosmic dust particles with complex composition impacting onto spacecraft materials.

Figure 4 shows the solution of Eq. (8) for the parameters of the experimental setup and those determined from the lab measurements described below. Due to the smaller physical size of the model, both the preshoot and the main peak are shifted to earlier times. The amplitudes of the impact signals are also reduced due to the limited capabilities of the dust accelerator ($Q_{IMP}$ is typically on the order of $10^{-14} - 10^{-15}$ C). The comparison of typical values between a SC in interplanetary space and the laboratory model are presented in Table. I.

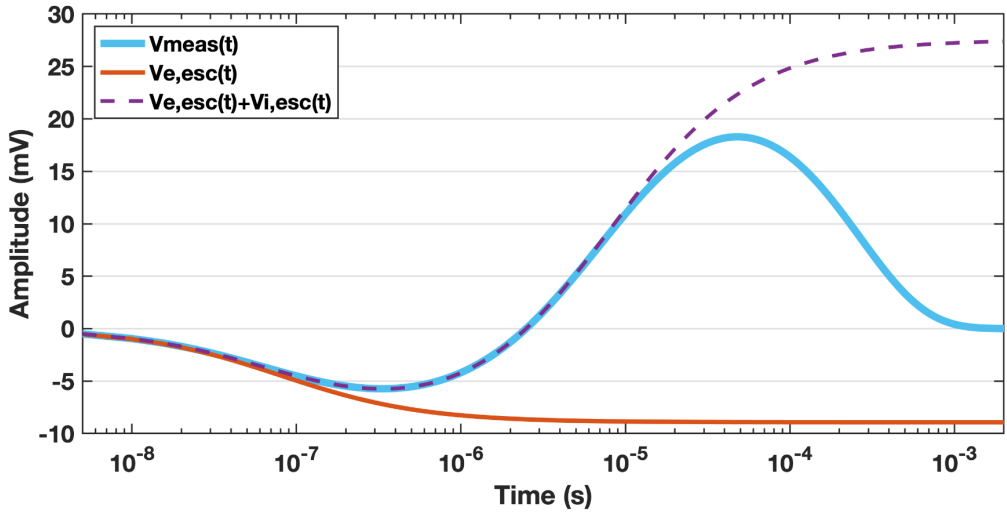

**Fig. 4:** Same as Fig. (3), except using the parameters of laboratory setup shown in Table. I.

**Table. I:** The list of parameters used for calculating the antenna signals shown in Figs. (3) and (4).

|  | Interplanetary space (Fig. 3) | Laboratory model (Fig. 4) |
|---|---|---|
| $V_{SC}$ | +5.5 V | +5.0 V |
| $Q_{IMP}$ | $2.0 \times 10^{-10} C$ | $3.5 \times 10^{-14} C$ |
| $C_{SC}$ | 200 pF | 48 pF |
| $R_{SC}$ | 1.0 m | 0.08 m |
| $v_e$ | 1,000 km/s | 1,000 km/s |
| $v_i$ | 10 km/s | 10 km/s |
| $T_e$ | 7 eV | 7 eV |
| $T_i$ | 25 eV | 25 eV |
| $\kappa$ | 0.50 | 0.50 |
| $\tau_{dis}$ | 53 µs (Eq. 7) | 240 µs = $R_{BIAS} C_{SC}$ |
| $\|Q_{e,esc}/Q_{IMP}\|$ | 0.23 | 0.24 |
| $\|Q_{e,col}/Q_{IMP}\|$ | 0.77 | 0.76 |

# 4. Analysis of the Experimental Data

A total of 182 dust impact waveforms were recorded using the setup described in Sec. 2. These are roughly equally distributed between the three applied bias voltage investigated (–5V, 0V, and +5V). Since the abundance of particles provided by the accelerator decreases with increasing speed [*Shu et al.*, 2012], most impact events occurred close to the 20 km/s lower limit, with the highest impact speed being about 40 km/s. The mass and velocity of the particles are measured in real time by detectors incorporated in the accelerator and allowed filtering for particles with speeds above 20 km/s. The typical mass of the impactors is about $10^{-18}$ kg.

The model presented in Sec. 3 is used to fit a subset of recorded waveforms and the fits provide the relevant parameters of the impact plasma. These include the capacitance of the SC ($C_{SC}$), the geometric coefficient ($\kappa$), the electron and ion expansion speeds ($v_e$ and $v_i$), their corresponding effective temperatures ($T_e$ and $T_i$), and the impact charge ($Q_{IMP}$).

## 4.1. Bias Potential $V_{SC} = 0$ V

Figure 5 shows the waveforms of three typical impact events for the SC bias voltage set to $V_{SC} = 0$ V. This condition allows examining the free expansion of the impact plasma from the impact location. The general shape of the signals is similar to those described in Sec. 3. The waveform starts with the negative-going preshoot due to the fast-escaping electrons. The expansion of electrons and ions from the impact plasma is over at around $t \approx 50$ μs. The subsequent exponential decay is the discharge of the net collected charge on the SC with a characteristic time constant of $\tau_{dis,L} = R_{BIAS} C_{SC}$. The first step in the data analysis is calculating the SC capacitance from this exponential decay, yielding the value of $C_{SC} = (48 \pm 8)$ pF. This capacitance value can be used to convert the measured voltage signals into charge. It is noted that the calculated value of $C_{SC}$ is lower than that found by *Nouzák et al.* [2018] by about a factor of 2.5 for a similar setup. The effective value of $C_{SC}$ is the sum from the physical capacitance of the SC body, and contributions from the parasitic capacitances to the mount, the antennas, the vacuum chamber, electronics boards, etc. The discrepancy between the values may be due to updated electronics boards, the slightly modified mounting brackets, or the routing of the cables, for example. The physical capacitance of the SC can also be estimated as $C = \varepsilon_0 \sqrt{4\pi S} \cong 18.6$ pF, where $S \cong 0.35$ m² is the surface area of the convex envelope around the model [*Chow and Yovanovich*, 1982].

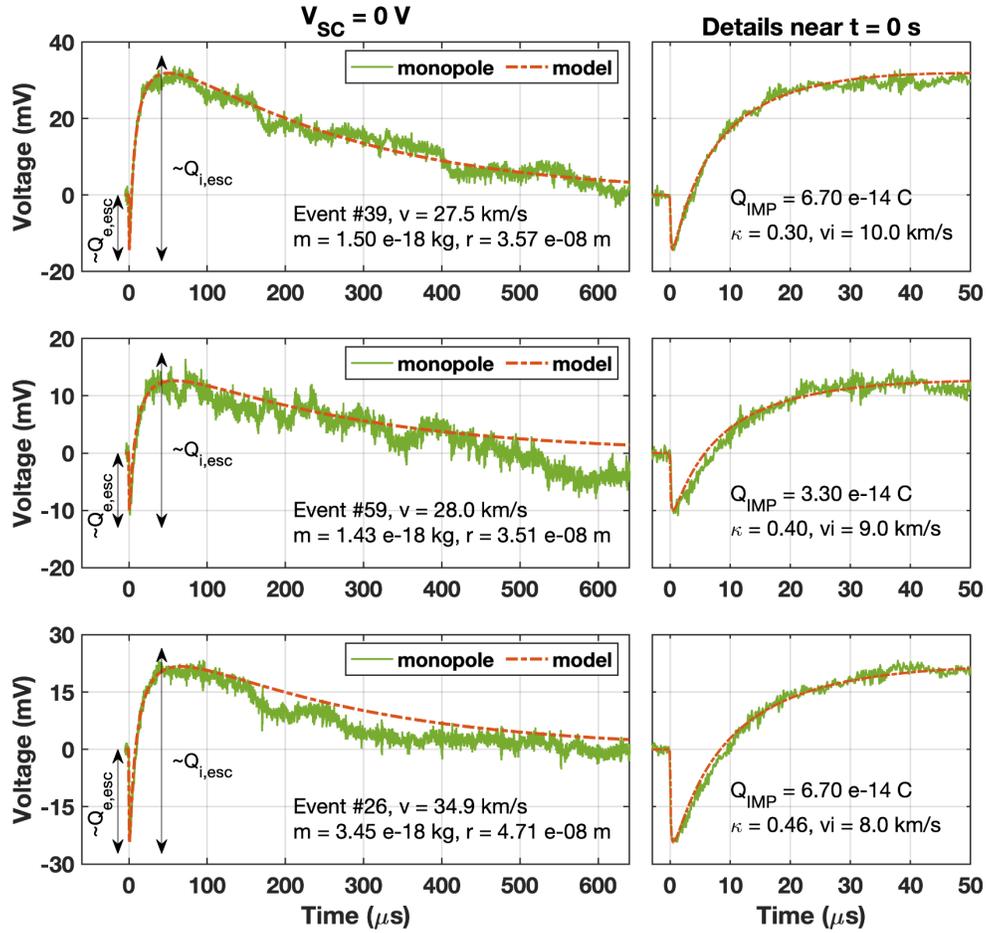

**Fig. 5:** Three example waveforms measured by the monopole antenna for $V_{SC} = 0$ V. The red dashed curves are from the model fitted to the data. The parameters of the particles are provided along with some of the parameters obtained from the fitting routine. The radius of the particle is provided as *r*. The approximate values of the escaping electron and ion charges are indicated. See text for more detail.

The next step in data analysis is finding the electron and ion expansion speeds. Following Eq. (6), the induced charge varies with time as $\sim 1/\left(1 + \frac{v_x t}{R_{SC}}\right)$, where $x$ refers to electrons or ions. The SC model, however, is not spherical, therefore, the 'effective' value of the SC radius is estimated to be $R_{SC} = 8$ cm (Sec. 2). For this dimension, the ion expansion speed varies between 5 – 15 km/s, with an average value of $v_i = (9.3 \pm 3.0)$ km/s. This ion expansion speed agrees well with values reported by *Collette et al.* [2016], *Kellogg et al.* [2016], and *Lee et al.* [2012]. The expansion of the much faster electrons generates the steep slope of the preshoot signal with a measured rise-time of about 0.4 μs that can be fully resolved by the updated fast electronics (Sec. 2). Nevertheless, finding a consistent value for $v_e$ turned out to be difficult, possibly due to the simultaneous expansion of both the electrons and ions, a wide energy distribution of the electrons, or large impact-to-impact variations. The electron escape speed is set to $v_e \cong 10^6$ m/s, as this value provides good fits to the data. This speed corresponds to the thermal speed of electrons with about 2 eV temperature. Calculations using the model suggest that by changing $v_e$ by a factor of 2, the change of the preshoot amplitude is limited to < 15%. The uncertainty associated with the selection of the value for $v_e$ is thus relatively small.

Figure 5 clearly indicates that the ratios of the escaping electron and ion charges are roughly equal to $Q_{e,esc}/Q_{i,esc} \approx 1/2$. Under the assumption that the number of electrons and ions generated by the impact is the same, this observation means that electrons are less likely to escape from the surface, even at zero bias potential. This observation is consistent with the physical picture of a stream of ions moving away from the impact location, while the faster electrons acquire an isotropic distribution. Once the expanding plasma cloud grows sufficiently large for the electrons and ions decouple from one another (typically within 1 mm from the surface in the lab measurements), half of the electrons are moving back towards the target, and half of the electrons away from the target. This physical picture is described in the model by the geometric coefficient $\kappa$ in Eqs. (3) and (4). Fitting the data for $V_{SC} = 0$ V bias voltage allows determining $\kappa$ independent of the electron temperature. The data analysis yielded a range of about $\kappa = 0.3 - 0.5$, with an average value of $\kappa = 0.405 \pm 0.078$. The fact that somewhat less than half of the electrons escape may be explained by the concave shape of the HGA, where the dust impacts occur (Sec. 2).

## 4.2. Bias Potential $V_{SC} = +5$ V

The waveforms recorded for a spacecraft bias potential $V_{SC} = +5$ V (Fig. 6) are generally similar to the zero-bias case. A notable difference is the reduced amplitude of the preshoot, as only a fraction of electrons is capable of overcoming the potential barrier from the applied positive bias voltage. The effective temperature of the electrons is then determined from the fitting of the data while keeping the $\kappa = 0.405$ parameter constant. The meaning of the effective temperature $T_e$ is defined by Eq. (3); noting that the expansion away from the surface of the SC is a 3D process, while the model is simplified to a spherically symmetric 1D case. The fitting of the data set yielded $T_e = (7.8 \pm 1.3)$ eV. This value is significantly higher than the ~1 eV value reported by *Collette* *et al.* [2016] and *Nouzák et al.* [2020] for the same impact speed range and dust-target material combination. The authors in the former source calculated $T_e$ from the statistical distribution of the retained charge on an impact plate as a function of the applied bias voltage. *Nouzák et al.* [2020] collected waveforms using a similar setup as presented in this article, and then statistically evaluated the reduction of the fraction of escaping electrons for 0 and +5 V bias potentials. In comparison, each waveform in this article has been fitted to the model from Sec. (3.1) in order to obtain a value for $T_e$. The measurements presented in this article are collected using electronics with an improved bandwidth that allowed fully resolving the preshoot part of the signals. This fact may at least partially describe the reason for the higher electron temperatures reported here. A recent study performed by *Kočiščák et al.* [2020] has measured the effective temperatures of the negative charge carriers in the impact plasma (assuming both free electrons and anions are present) for an olivine dust sample and tungsten target. The particles were coated with a conductive polymer in this study that enabled their acceleration. The temperature reported for an impact velocity range of 12–18 km/s was 4.4 eV, which is closer to the value reported here.

The routine used for fitting the model to the measured waveforms treats the electron expansion speed ($v_e$) and electron temperature ($T_e$) as independent variables, while in reality they are naturally related. The selected expansion speed of $v_e = 10^6$ m/s corresponds to about 2 eV electron temperature, which is significantly lower than the 7.8 eV effective temperature determined above. This is not necessarily a contradiction, however, as in the assumed isotropic expansion of the electrons, the radial component of the velocity varies over a wide range depending on the elevation angle of the electrons with respect to the surface normal at the location of the impact. The $v_e$ parameter thus describes the effective radial expansion speed of the electrons over their entire distribution. As discussed in Sec. (3), the amplitude of the preshoot is also affected by the ratio of the electron and ion expansion speeds.

The ion expansion speed determined for the $V_{SC} = +5$ V case is $v_i = (9.3 \pm 2.3)$ km/s, i.e., essentially the same as the zero-bias case. This makes sense, as the applied bias is small in comparison to the kinetic energy of the cations. Besides, the potential profile scales with a distance $d$ from the surface of the SC as $\sim 1/(R_{SC} + d)$, and the ion speed is only weakly affected in the early phases of the expansion.

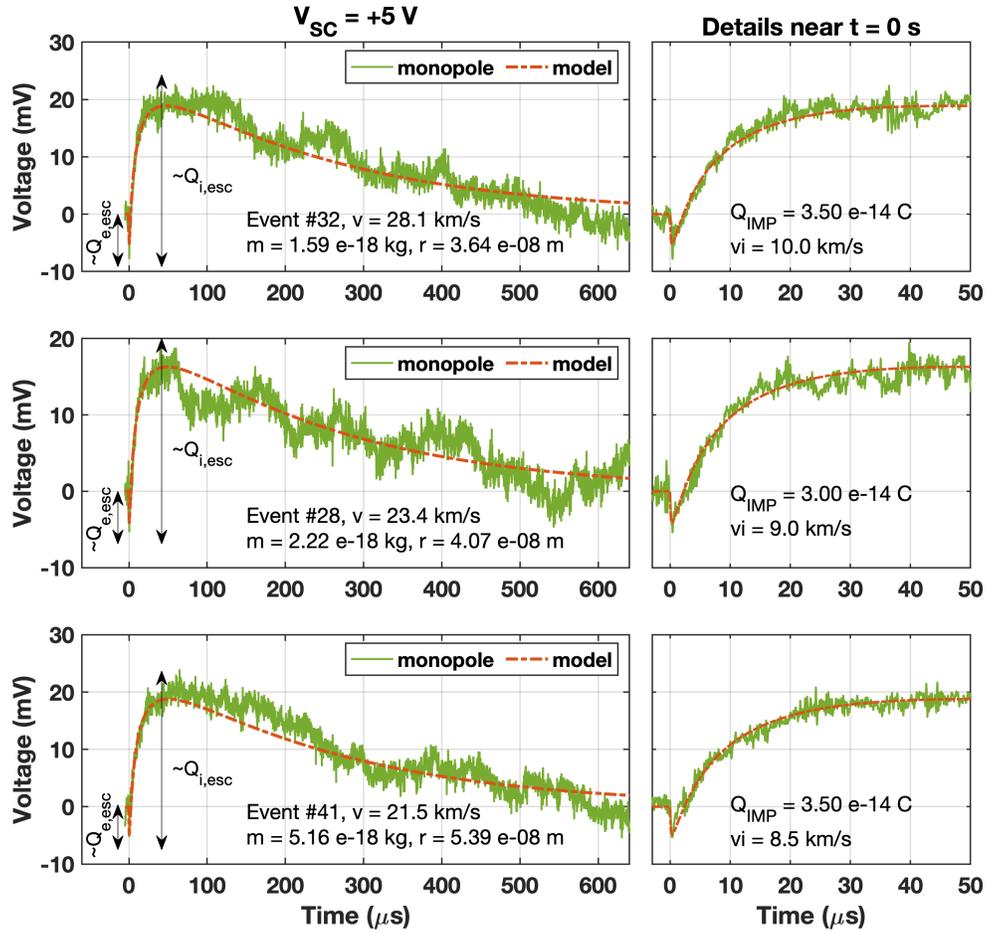

**Fig. 6:** Three example waveforms measured by the monopole antenna for $V_{SC} = 5$ V. The labeling is the same as in Fig. 5.

## 4.3. Bias potential $V_{SC} = -5$ V

Figure 7 shows waveforms for negative bias potential $V_{SC} = -5$ V, where the preshoot component is again pronounced. Following the argument from the bottom of the previous section, the negative bias has little effect on the collected vs. escaping electron fraction, and thus $\kappa = 0.405$ is assumed. The effective temperature of the ions can be calculated similarly to those of the electrons using Eq. (4). The result is $T_i = (25.4 \pm 12.5)$ eV, noting that the confidence in this number is lower, given that the ion temperature is high compared to the applied bias potential. Nevertheless, it is pointed out that the effective ion temperature for this impact speed range is in good agreement with 23 eV value measured by *Collette et al.* [2016], while the value determined by *Nouzák et al.* [2020] was 10–15 eV. The cation temperature measured for an impact speed range of 12–18 km/s and polymer-coated olivine particles was approximately 7 eV [*Kočiščák et al.*, 2020]. The ion expansion speed is $v_i = (10.0 \pm 3.1)$ km/s that is similar to the two cases from above. Since the ion temperature is significantly larger than the bias potential, the ion expansion speed is not significantly affected. It is interesting to note that the expansion speed and ion temperature values are consistent with one another for an ion mass of Fe ($m = 55.8$ u). Iron ions are no doubt one of the most abundant cation species in the impact plasma at these impact velocities. There is one noticeable deviation in the waveforms collected for negative SC bias, when compared to the other two cases. The preshoot peak is wider at the bottom and there is a ledge and a change in the slope (Fig. 7). This feature is generated as an artifact by the front-end amplifier (AD8421 Instrumentation Amplifier).

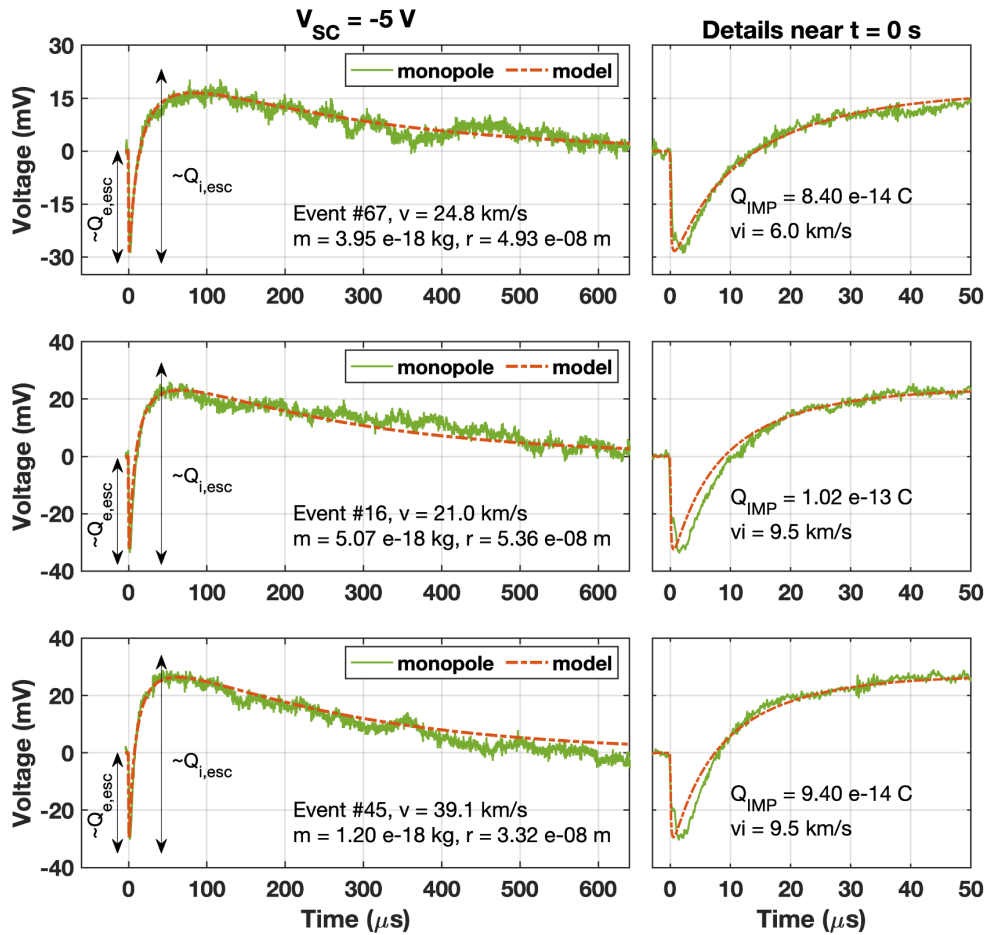

**Fig. 7:** Three example waveforms measured by the monopole antenna for $V_{SC} = -5$ V. The labeling is the same as in Fig. 5.

## 4.4. Impact Charge $Q_{IMP}$

The impact charge, $Q_{IMP}$, is determined for each analyzed waveform. The impact charge is an important parameter as it is characteristic to the mass and speed of the dust particle, according to Eq. (1). The amplitude of the main peak of the waveform is related to $Q_{IMP}$, but it is also affected by the effective temperatures of the positive and negative charge carriers, the geometric coefficient ($\kappa$), and mostly by the discharge time constant, as demonstrated in Sec. 3.

Figure 8 shows $Q_{IMP}/m$ as a function of impact speed for all waveforms analyzed in detail. Several observations are ought to be made here. First, ratio scales with velocity as $\sim v^{3.7}$, which is consistent with the $\sim v^{3.5}$ scaling from prior measurements over a wider velocity range [*Dietzel et al.,* 1973]. The absolute value of the charge is about a factor of two lower, however. This may be explained by neglecting the coupling parameter $\Gamma$ from Sec. 3.1. Since the potential developed on the SC from impact charge is also affecting the potential of the antenna through capacitive coupling, the measured signal $V_{meas}(t)$ is effectively reduced. The relatively large scatter of the $Q_{IMP}/m$ ratio is characteristic to impact ionization process [e.g., *Horanyi et al.,* 2014].

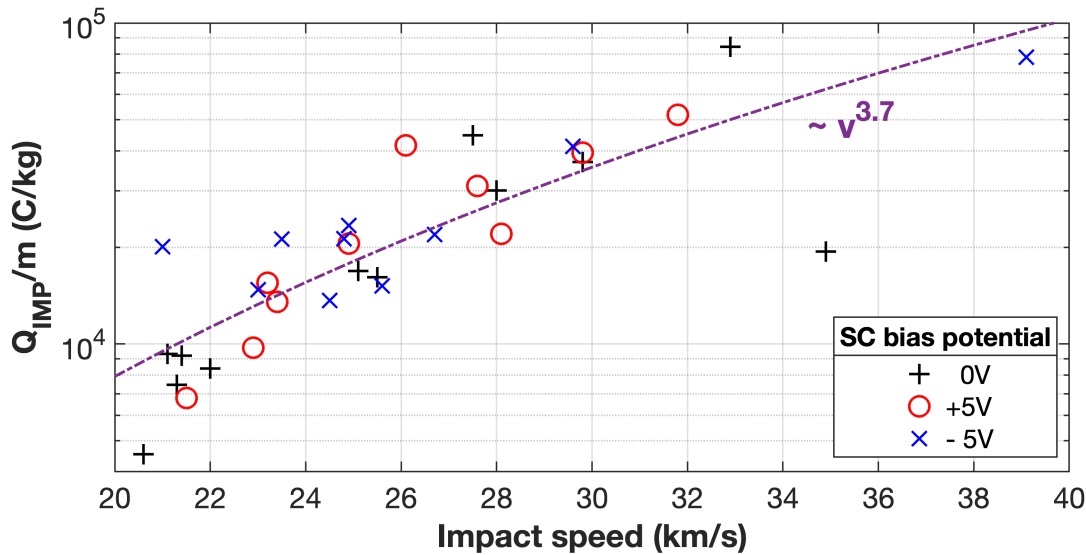

**Fig. 8:** The normalized impact charge ($Q_{IMP}/m$) as a function of impact speed calculated for the three different SC bias potentials. See text for more details.

## 5. Discussion and Summary

The main result from the work presented above is that there is good agreement between the presented antenna signal model and the data collected from laboratory measurements using a reduced-size model and the dust accelerator facility. Moreover, the parameters derived from fitting the model to the measurements reasonably agree with what is expected based on prior studies. In summary, the four key elements of this model are the following: First, the impact charge can be subdivided into fractions of escaping and collected charges for both the negative (electrons) and positive (cation) components. Second, the escaping/collected fractions of the impact cloud are determined by the energetics of the system and the nature of the impact plasma cloud. The energetics means that escaping/collected fractions are determined by the potential of the spacecraft and the effective temperatures of the charged species. Furthermore, the presented work implies that it is reasonable to think about the impact plasma as a plume of ions expanding away from the impact location while the electrons acquire an isotropic distribution during the expansion. Third, the voltages developing on the antenna and spacecraft elements can be understood based on electrostatics. This means that the direct charging due to the collected fractions from the impact plasma, and the induced charging due to the escaping fractions of the impact plasma are identified as the main mechanisms for generating the antenna signals. And fourth, the restoration driven by photoelectron emission and the ambient plasma environment to equilibrium potential needs to be considered. The corresponding discharging currents may and likely will significantly limit the amplitude of the signal, and thus detailed analysis is needed in order to convert the antenna signals to impact charge, which is a measure of the dust particles mass if impact speed is known.

Several simplifying assumptions have been used for deriving the antenna signal model. The two most important ones are the limited impact charge and impact location being far from the antenna. Large dust particles impacting at high speed may generate massive impact charges for which the presented model will no longer be valid. The latter assumption implies that the impact charge cloud does not interact with the antenna. If this assumption is not satisfied, the direct or induced charging of the antenna needs to be included in the model, as the voltage on the antenna will also deviate from equilibrium. For completeness, it is also noted that the present model does not provide a quantitative description of how antennas operated in a dipole mode detect dust impact events.

As a last comment, it is noted that while the model has been evaluated for high impact speed ($\geq$ 20 km/s), there is no reason why it could not be expanded to lower speeds. The high impact speeds investigated provide impact clouds with the 'simplest' parameters. These include the instantaneous generation of the impact plasma and electrons being the dominant carriers of the negative charges. At lower speeds, the impact plasma is generated over a finite time interval, which is on the order of about 10 μs [e.g., *Collette et al.*, 2013]. Moreover, the composition of the impact plasma, including the electron/anion ratio, may change rapidly with impact speed. With that said, the physical model of the antenna signal generation has to be the same once the aforementioned effects are included. As a future investigation, it may be possible to use the presented model to infer the properties of the impact plasma at low impact speeds from measured laboratory antenna signals.

**Acknowledgement**

The contributions to this study from authors M.S., M.H., H.H., and Z.S. were supported by NASA's Cassini Data Analysis Program (CDAP), Grant NNX17AF99G. The contribution from author D.M. was supported through a dust debris damage assessment augmentation to NASA contract NNN06AA01C. The FIELDS experiment on Parker Solar Probe was designed and developed under this contract. The operation of the dust accelerator was supported by NASA's Solar System Exploration Research Virtual Institute (SSERVI) Cooperative Agreement Notice, Grant 80NSSC19M0217. The authors thank David James and John Fontanese for operating the dust accelerator during the experimental campaigns. The laboratory data (Shen et al., 2021) are publicly available in Zenodo repository (https://doi.org/10.5281/zenodo.4285807).

Manuscript accepted online by JGR: Space Physics on 05 April 2021.   https://doi.org/10.1029/2020JA028965# References

Aubier, M.G., Meyer-Vernet, N. and Pedersen, B.M. (1983), Shot noise from grain and particle impacts in Saturn's ring plane. Geophys. Res. Lett., 10: 5-8. https://doi.org/10.1029/GL010i001p00005

Auer, S., (2001). Instrumentation, in Interplanetary Dust, edited by E. Grün, pp. 385–444, Springer, New York.

Chow, Y. L., and M. M. Yovanovich (1982). The Shape Factor of the Capacitance of a Conductor. *Journal of Applied Physics*, *53*(12), 8470–8475.

Collette, A., Drake, K., Mocker, A., Sternovsky, Z., Munsat, T., & Horányi, M. (2013). Time-resolved temperature measurements in hypervelocity dust impact. *Planetary and Space Science*, *89*, 58–62. http://doi.org/10.1016/j.pss.2013.02.007

Collette, A., Grün, E., Malaspina, D., and Sternovsky, Z. (2014), Micrometeoroid impact charge yield for common spacecraft materials, J. Geophys. Res. Space Physics, 119, 6019–6026, doi:10.1002/2014JA020042.

Collette, A., Meyer, G., Malaspina, D., and Sternovsky, Z., 2015. Laboratory investigation of antenna signals from dust impacts on spacecraft. J. Geophys. Res.: Space Physics 120 (7), 5298–5305.

Collette, A., Malaspina, D., and Sternovsky, Z., (2016). Characteristic temperatures of hypervelocity dust impact plasmas. J. Geophys. Res.: Space Physics 121 (9), 8182–8187.

Dietzel, H., Eichhorn, G., Fechtig, H., Grun, E., Hoffmann, H.-J., & Kissel, J. (1973). The HEOS 2 and HELIOS micrometeoroid experiments. Journal of Physics E: Scientific Instruments, 6(3), 209–217. https://doi.org/10.1088/0022-3735/6/3/008

Fielding, L. A., Hillier, J. K., Burchell, M. J., & Armes, S. P. (2015). Space science applications for conducting polymer particles: synthetic mimics for cosmic dust and micrometeorites. *Chemical Communications*, *51*(95), 16886–16899. http://doi.org/10.1039/C5CC07405C.

Gurnett, D. A., Grün, E., Gallagher, D., Kurth, W. S., & Scarf, F. L. (1983). Micron-sized particles detected near Saturn by the Voyager plasma wave instrument. Icarus, 53(2), 236–254. https://doi.org/10.1016/0019-1035(83)90145-8

Gurnett, D. A., Ansher, J. A., Kurth, W. S., & Granroth, L. J. (1997). Micron-sized dust particles detected in the outer solar system by the Voyager 1 and 2 plasma wave instruments. Geophysical Research Letters, 24(24), 3125–3128. https://doi.org/10.1029/97gl03228

Gurnett, D. A., Kurth, W. S., Kirchner, D. L., Hospodarsky, G. B., Averkamp, T. F. et al., 2004. The Cassini Radio and Plasma Wave investigation. Space Science Reviews 114 (1–4), 395–463.

Gurnett, D. A., Kurth, W. S., Hospodarsky, G. B., Persoon, A. M., Averkamp, T. F., Cecconi, B., ... & Galopeau, P. (2005). Radio and plasma wave observations at Saturn from Cassini's approach and first orbit. Science, 307(5713), 1255-1259.

Hill, T. W., Thomsen, M. F., Toka, R. L., Coates, A. J., Lewis, G. R., Young, D. T., Crary, F. J., Baragiola, R. A., Johnson, R. E., Dong, Y., Wilson, R. J., Jones, G. H., Wahlund, J.-E., Mitchell, D. G. and Horányi, M., 2012. Charged nanograins in the Enceladus plume. J. Geophys. Res.: Space Physics 117 (A05209).

Hillier, Jon K., et al. "Impact ionisation mass spectrometry of polypyrrole-coated pyrrhotite microparticles." Planetary and Space Science 97 (2014): 9-22.
24